\documentclass[a4paper,11pt]{article}
\pdfoutput=1 

\usepackage{jcappub} 
\usepackage[T1]{fontenc} 
\usepackage{tikz-feynhand}
\usepackage{graphicx}
\usepackage{xcolor}
\usepackage{caption,subcaption}
\usepackage{mathrsfs,mathtools}
\usepackage{physics,amssymb}
\usepackage{bm}
\usepackage{braket}
\usepackage{listings}
\usepackage{cases}
\usepackage{comment}
\usepackage{soul}
\usepackage{cancel}
\usepackage{cases}
\usepackage[utf8]{inputenc}
\usepackage{url}
\usepackage[normalem]{ulem}
\usepackage{xspace}
\usepackage{aas_macros}
\usepackage{acronym}
\usepackage{siunitx}

\hypersetup{colorlinks=true
,urlcolor=DARKBLUE
,anchorcolor=DARKBLUE
,citecolor=DARKBLUE
,filecolor=DARKBLUE
,linkcolor=DARKBLUE
,menucolor=DARKBLUE
,linktocpage=true
,pdfproducer=medialab
,pdfa=true
}

\definecolor{MONZA}{HTML}{CF000F}
\definecolor{DARKBLUE}{HTML}{00008b}
\definecolor{DARKMAGENTA}{HTML}{8b008b}

\newcommand{\HH}{\mathcal{H}}
\newcommand{\PP}{\mathcal{P}}

\newcommand{\ee}{\mathrm{e}}

\newcommand{\uth}{\mathrm{th}}
\newcommand{\PBH}{\mathrm{PBH}}
\newcommand{\DM}{\mathrm{DM}}
\newcommand{\tot}{\mathrm{tot}}
\newcommand{\GW}{\mathrm{GW}}
\newcommand{\RD}{\mathrm{RD}}

\newcommand{\dk}{\frac{\dd^3k}{(2\pi)^3}}
\newcommand{\iq}[1]{\int\frac{\dd^3q_{#1}}{(2\pi)^3}}

\newcommand{\hk}[2]{h^{#2}_{\lambda^{#1}}(\tau,\bfk^{#1})}
\newcommand{\Q}[2]{Q_{\lambda_{#1}}(\bfk_{#1},\bfq_{#2})}

\acrodef{CMB}{cosmic microwave background}
\acrodef{PBH}{primordial black hole}
\acrodef{PDF}{probability density function}
\acrodef{EoM}{equation of motion}
\acrodef{GW}{gravitational wave}
\acrodef{RD}{radiation-dominated}
\acrodef{DM}{dark matter}
\acrodef{PTA}{pulsar timing array}
\acrodef{SIGW}{scalar induced gravitational wave}
\acrodef{SGWB}{stochastic gravitational wave background}
\acrodef{BSSN}{Baumgarte--Shapiro--Shibata--Nakamura}
\acrodef{LISA}{Laser Interferometer Space Antenna}
\acrodef{NANOGrav}{North American Nanohertz Observatory for Gravitational Waves}
\acrodef{DECIGO}{Decihertz Gravitational Wave Observatory}
\acrodef{FLRW}{Friedman–Lemaitre–Robertson–Walker}

\newcommand{\uc}{\mathrm{c}}
\newcommand{\calC}{\mathcal{C}}

\newcommand{\calH}{\mathcal{H}}

\newcommand{\bfk}{\mathbf{k}}

\newcommand{\um}{\mathrm{m}}
\newcommand{\calN}{\mathcal{N}}
\newcommand{\calO}{\mathcal{O}}

\newcommand{\calP}{\mathcal{P}}

\newcommand{\bfq}{\mathbf{q}}

\newcommand{\ur}{\mathrm{r}}

\newcommand{\bfx}{\mathbf{x}}

\newcommand{\bae}[1]{\begin{align} #1 \end{align}}
\newcommand{\beae}[1]{\begin{equation}\begin{aligned} #1 \end{aligned}\end{equation}}

\newcommand{\bme}[1]{\begin{multline} #1 \end{multline}}

\newcommand{\lr}[1]{\left( #1 \right)}
\newcommand{\bce}[1]{\begin{cases} #1 \end{cases}}

\newcommand{\hD}{\hat{\Delta}}

\newcommand{\zero}{{_0}}
\newcommand{\munu}{_{\mu\nu}}

\newcommand{\rhofl}{\rho_{\rm fl}}
\newcommand{\rhozero}{\rho_{\zero}}
\newcommand{\be}{\begin{equation}}
\newcommand{\beqn}{\begin{eqnarray}}
\newcommand{\eeq}{\end{equation}}
\newcommand{\eeqn}{\end{eqnarray}}

\begin{document}

\author[a]{Ryoto Inui,}
\author[b]{Cristian Joana,}
\author[c]{Hayato Motohashi,}
\author[b,d,e]{Shi Pi,}
\author[a,f]{Yuichiro Tada,}
\author[g,a,e]{and Shuichiro Yokoyama}


\affiliation[a]{Department of Physics, Nagoya University, 
Furo-cho Chikusa-ku,
Nagoya 464-8602, Japan}
\affiliation[b]{CAS Key Laboratory of Theoretical Physics, Institute of Theoretical Physics,
Chinese Academy of Sciences, Beijing 100190, China}
\affiliation[c]{Division of Liberal Arts, Kogakuin University, 2665-1 Nakano-machi, Hachioji, Tokyo, 192-0015, Japan}
\affiliation[d]{Center for High Energy Physics, Peking University, Beijing 100871, China}
\affiliation[e]{Kavli Institute for the Physics and Mathematics of the Universe (WPI), UTIAS, The University of Tokyo, Kashiwa, Chiba 277-8583, Japan}
\affiliation[f]{Institute for Advanced Research, Nagoya University,
Furo-cho Chikusa-ku, 
Nagoya 464-8601, Japan}
\affiliation[g]{Kobayashi Maskawa Institute, Nagoya University, 
Chikusa, Aichi 464-8602, Japan}

\emailAdd{inui.ryoto.a3@s.mail.nagoya-u.ac.jp}
\emailAdd{cristian.joana@itp.ac.cn}
\emailAdd{motohashi@cc.kogakuin.ac.jp}
\emailAdd{shi.pi@itp.ac.cn}
\emailAdd{tada.yuichiro.y8@f.mail.nagoya-u.ac.jp}
\emailAdd{shu@kmi.nagoya-u.ac.jp}

\title{\boldmath Primordial black holes \\
and induced gravitational waves from logarithmic non-Gaussianity}
\date{\today}

\abstract{We investigate the formation of \ac{PBH} based on numerical relativity simulations and peak theory  
as well as the corresponding \ac{SIGW} signals in the presence of \emph{logarithmic non-Gaussianities} which has recently been confirmed in a wide class of inflation models.
Through numerical calculations, we find certain parameter spaces of the critical thresholds for the type A \ac{PBH} formation and reveal a maximum critical threshold value. We also find that there is a region where no \ac{PBH} is produced from type II fluctuations contrary to a previous study. 
We then confirm that \ac{SIGW} signals originated from the logarithmic non-Gaussianity are detectable in the Laser Interferometer Space Antenna 
if \acp{PBH} account for whole dark matter. Finally, we discuss the \ac{SIGW} interpretation of the nHz stochastic gravitational wave background reported by the recent pulsar timing array observations. We find that PBH overproduction is a serious problem for most of the parameter space, while this tension might still be alleviated in the non-perturbative regime.}

\maketitle
\flushbottom

\acresetall
\section{Introduction}
\label{sec:intro}
\Acp{PBH} are hypothetical compact objects 
that might have been formed in the early universe~\cite{Hawking:1971ei,Carr:1974nx}. Although various formation scenarios have been discussed (e.g., the formation in the matter-dominated universe~\cite{Harada:2016mhb}, the formation from the isocurvature perturbation~\cite{Passaglia:2021jla,Yoo:2021fxs}, and the formation from a resonant instability of cosmological perturbation during a preheating~\cite{Khlopov:1985fch, Martin:2019nuw}), the widely discussed one is the gravitational collapse of the highly dense regions in the radiation-dominated universe. \Acp{PBH} can be formed in a wide range of masses unlike astrophysical black holes, and they can provide a sizable amount of \ac{DM}.  
The mass range between $[10^{-15}M_{\odot}\text{--}10^{-11}M_{\odot}]$ is often called the \emph{\ac{PBH} Dark Matter window} because it can account for all the \ac{DM} according to the current observational constraints~\cite{Carr:2020xqk,Green:2020jor,Carr:2020gox}. \Acp{PBH} are not only promising \ac{DM} candidates but can also potentially explain other cosmological and astrophysical phenomena. For instance, they could constitute the seeds for the supermassive black holes in galactic nuclei or galaxies~\cite{Carr:2018rid}, be relevant sources of \ac{GW} events in ground-based detectors~\cite{Bird:2016dcv, Sasaki:2016jop, Clesse:2016vqa, Huang:2024wse} and in \ac{PTA} observations~\cite{NANOGrav:2023hvm, EPTA:2023fyk, Reardon:2023gzh, Xu:2023wog}, have a role on explaining baryogenesis~\cite{Carr:2019hud,Garcia-Bellido:2019vlf,Despontin:2024jzp}, and even stand as the exotic object present in the solar system~\cite{Witten:2020ifl,Domenech:2020ers}. 

In addition to these motivations, recently, more precise studies focusing on the relationship between \acp{GW} and \acp{PBH} have been actively discussed (see, e.g., Ref.~\cite{Domenech:2021ztg,Domenech:2023fuz}). The large primordial perturbations necessary for \ac{PBH} formation can induce \acp{GW} through the second-order interactions between the scalar and tensor metric perturbations. The frequency of \acp{GW} is related to the \ac{PBH} mass, as both of them depend on the size of the Hubble horizon at the reentry. In particular, our main interest relies on the mass band of the \ac{PBH} mass window corresponding to the frequency band of space-borne interferometers, such as \ac{LISA}~\cite{LISA:2017pwj}, \ac{DECIGO}~\cite{Kawamura:2011zz}, Taiji~\cite{Ruan:2018tsw}, and TianQin~\cite{TianQin:2015yph}. In fact, it was pointed out that \ac{LISA} can detect the \ac{SIGW} signal when the \acp{PBH} in the mass range of the \ac{DM} window occupy the entire \ac{DM}~\cite{Cai:2018dig,Bartolo:2018evs}. Interestingly, the detectability of milihertz \ac{SIGW} is robust against non-Gaussianity, which has also been widely studied~\cite{Cai:2018dig,Unal:2018yaa, Adshead:2021hnm, Abe:2022xur,Pi:2024jwt, He:2024luf, Papanikolaou:2024kjb}. 

On the theoretical side, recently, the curvature perturbations $\zeta$ having a non-Gaussian exponential tail in their probability density function have come to be discussed~\cite{Pattison:2017mbe,Ezquiaga:2019ftu,Figueroa:2020jkf,Pattison:2021oen}. One typical model that realizes such a non-Gaussian curvature perturbation is the ultra slow-roll inflation. 
The primordial curvature perturbation produced during the ultra slow-roll phase could have sufficiently large amplitudes for \ac{PBH} formations and have an \emph{exponential-tail} distribution 
which asymptotically follows $P(\zeta) \propto \exp(-3\zeta)$ in the large $\zeta$ limit. 
It is explained by the logarithmic relation $\zeta = -(1/3) \ln{(1 - 3 \zeta_g)}$ between the curvature perturbation $\zeta$ and a certain Gaussian random field $\zeta_g$.
A more general form of a logarithmic type curvature perturbation can be expressed as $\zeta = -(1/\gamma) \ln{(1 - \gamma \zeta_g)}$ with a constant parameter $\gamma$, which is smoothly connected to the Gaussian fluctuation $\zeta_g$ for $\gamma \rightarrow 0$.  We call this the \emph{logarithmic non-Gaussianity} relation. This logarithmic non-Gaussianity can be found in a class of the constant roll inflation model which is the generalized model of the ultra slow-roll inflation~\cite{Atal:2019cdz, Atal:2019erb, Pi:2022ysn, Wang:2024xdl, Inui:2024sce}, as well as in the curvaton scenario \cite{Pi:2021dft}. 
In the constant roll scenario, the primordial scalar power spectrum can be blue-tilted and it could realize the efficient \ac{PBH} formation~\cite{Motohashi:2014ppa, Motohashi:2017aob, Motohashi:2019tyj, Motohashi:2017vdc, Motohashi:2019rhu, Tomberg:2023kli, Motohashi:2023syh}. 
The probability density function of the logarithmic type curvature perturbations has the exponential tail $P(\zeta)\propto \exp(-\gamma \zeta)$ for $\gamma > 0$, and follows the \emph{Gumbel} tail as $P(\zeta)\propto \exp\left(-\frac{1}{2\gamma^2\sigma^2}\ee^{-2\gamma\zeta}\right)$ with the variance $\sigma^2 = \braket{\zeta^2_g} $ for $-3/2 < \gamma < 0$~\cite{Pi:2022ysn,Inui:2024sce}.
Although the influence of the exponential tail in the ultra slow-roll model (i.e., $\gamma =3$) 
on the \ac{PBH} formation and the corresponding \ac{SIGW} signals have been studied in Refs.~\cite{Kitajima:2021fpq,Abe:2022xur}, the \ac{PBH} formation and the related \ac{SIGW} signals in general parameter space are unknown. Additionally, as PBHs may be overproduced when trying to explain the nHz SGWB observed by PTAs as induced by the curvature perturbation \cite{NANOGrav:2023hvm}, a negative non-Gaussianity is implied \cite{Ferrante:2022mui}. This also motivates us to study the PBH formation in the logarithmic non-Gaussianity with $\gamma<0$.

In this work, we investigate the \ac{PBH} formation when there is logarithmic non-Gaussianity, based on the numerical relativity and peak theory. The corresponding \ac{SIGW} spectrum is also calculated analytically. This paper is organized as follows. We describe the process of PBH formation including the considered peak profile, threshold estimation, and computation of abundances in Sec.~\ref{sec: peak_prof}. Next, in Sec. \ref{sec: SIGW}, we briefly review the perturbative formula for the \ac{SIGW} with the primordial scalar non-Gaussianities based on Ref.~\cite{Abe:2022xur} and provide the predicted signal in relation to current and future experiments. We investigate the detectability for the \ac{SIGW} with the logarithmic non-Gaussianity in light of \ac{PBH} \ac{DM} scenario by the \ac{LISA}. We also perform the parameter estimation for \ac{SIGW} in light of the logarithmic non-Gaussianity based on the recent \ac{PTA} results on the stochastic gravitational wave background, and discuss the compatibility of the abundance of \acp{PBH}.
Finally, we provide our conclusions in Sec.~\ref{sec: conclusion}.

\section{\ac{PBH} formation with the logarithmic non-Gaussianity}
\label{sec: peak_prof}

In this section, we investigate the \ac{PBH} formation from the primordial curvature perturbation with the logarithmic non-Gaussianity based on peak theory. 
In general, primordial curvature perturbations can have non-Gaussianities when there is nonlinear evolution. Among several types of non-Gaussianities, in this work, we 
focus on a special local-type non-Gaussianity, of which the primordial curvature perturbations $\zeta$ can be expressed as a logarithmic function of the Gaussian fluctuations $\zeta_g$ as
\bae{\label{general_expotail}
\zeta({\bf x}) = -\frac{1}{\gamma}\ln(1 - \gamma \zeta_g({\bf x})),
}
where $\gamma$ is a parameter that controls the tail behavior of the probability density distribution of the curvature perturbation. 
Such a logarithmic type of primordial curvature perturbation appears in several inflationary models which can produce \acp{PBH} efficiently, \textit{e.g.}~\cite{Atal:2019cdz, Atal:2019erb, Pi:2021dft, Pi:2022ysn, Wang:2024xdl, Inui:2024sce}. While the ultra-slow-roll model gives $\gamma = 3$ as mentioned in Sec.~\ref{sec:intro}, the parameter $\gamma$ can be also related to the inflaton mass (curvature of the inflaton potential) $\eta_{V} = V^{\prime\prime} / V$ in specific models. For example, in the quadratic constant-roll model, the relation between the inflaton mass and the $\gamma$ is given by Eq.~(3.8) in Ref.~\cite{Inui:2024sce} (see also Ref.~\cite{Pi:2022ysn}).
The statistical properties of the curvature perturbation differ depending on whether $\gamma$ is positive or negative. 
The probability distribution of $\zeta$ can be calculated as \cite{Pi:2022ysn}
\begin{align}
    P(\zeta) 
    = \abs{ \dv{\zeta_g}{\zeta} } P_g\qty(\zeta_g(\zeta))  = \frac{1}{\sqrt{2 \pi \sigma^2}} \exp 
    \left[ - \frac{1}{2 \gamma^2 \sigma^2} \left( \ee^{-\gamma \zeta} - 1 \right)^2 - \gamma \zeta
    \right]~,
\end{align}
where $P_{g}(\zeta_g)$ is the Gaussian distribution and $\sigma^2 = \Braket{\zeta_g^2}$ denotes the variance of $\zeta_g$.
The tail behavior of the probability density function of $\zeta$ is proportional to $\exp(-\gamma \zeta)$ for $\gamma >0$ and $\exp\left(- \frac{1}{2 \gamma^2 \sigma^2} \ee^{-2\gamma \zeta}\right)$ for $\gamma < 0$  respectively. In the probability density distribution, the signature of $\gamma$ reflects the sign of skewness, which is very important for the PBH abundance. While $\gamma\to0$ limit is the Gaussian case, $|\gamma|\ll1$ is the standard slow-roll inflation, of which the higher orders can be safely neglected. $\gamma>1$ (or more generally $|\gamma|>1$) implies the higher order terms can not be neglected, i.e., the non-Gaussianity is strong, thus the perturbative series does not work.

\subsection{Peak profile}
 
The radial profile of the peak of the curvature perturbation $\hat{\zeta}(r)$ is also related to the peak profile of the Gaussian fluctuation $\hat{\zeta}_g(r)$ by the logarithmic mapping~\cite{Yoo:2019pma, Kitajima:2021fpq} 
\bae{
\label{eq: prof_general}
\hat{\zeta}(r) = -\frac{1}{\gamma}\ln(1 - \gamma \hat{\zeta}_g(r)).
} 
We here suppose that the typical peak profile is spherically symmetric about the peak position $\bfx=\mathbf{0}$ as suggested by the peak theory~\cite{Bardeen:1985tr} we will introduce later.\footnote{See Ref.~\cite{Escriva:2024aeo} for the effect of non-sphericity.} The peak profile of a Gaussian field is given by~\cite{Yoo:2018kvb, Yoo:2020dkz, Kitajima:2021fpq},
\bae{\label{zetag_prof}
    \hat{\zeta}_g(r)=\mu_2\bqty{\frac{1}{1-\gamma_3^2}\pqty{\psi_1(r)+\frac{1}{3}R_3^2\Delta\psi_1(r)}-\frac{\mu_3^2}{\gamma_3(1-\gamma_3^2)}\pqty{\gamma_3^2\psi_1(r)+\frac{1}{3}R_3^2\Delta\psi_1(r)}}+\zeta_g^\infty,
}
which is characterized by three parameters: the peak height $\mu_2$, the width $\mu_3$, and the overall offset  $\zeta_g^\infty(= \hat{\zeta}_g \big|_{r \rightarrow \infty})$.
The statistical quantities and the two-point correlation function are given by
\beae{\label{stat_param}
    \sigma_n^2&=\int\frac{\dd{k}}{k}k^{2n}\calP_g(k), \quad & \psi_n(r)&=\frac{1}{\sigma_n^2}\int\frac{\dd{k}}{k}k^{2n}\frac{\sin(kr)}{kr}\calP_g(k), \\
    \gamma_n&=\frac{\sigma_n^2}{\sigma_{n-1}\sigma_{n+1}}, \quad & R_n&=\frac{\sqrt{3}\sigma_n}{\sigma_{n+1}}, 
    \quad \text{for odd $n$},
}
with $\zeta_g$'s power spectrum
\bae{
    \calP_g(k)=\frac{k^3}{2\pi^2}\int\dd[3]{x}\ee^{-i\bfk\cdot\bfx}\Braket{\zeta_g\qty(\frac{\bfx}{2})\zeta_g\qty(-\frac{\bfx}{2})}.
}
We hereafter assume a monochromatic power spectrum for $\zeta_g$, with amplitude $A_g$, which has a Dirac delta peak at $k=k_*$, 
\bae{\label{eq: monochromatic P}
    \calP_g(k)=A_g\delta(\ln k-\ln k_*)~.
}
In such a case, the peak profile of Eq.~(\ref{zetag_prof}) reduces to the simpler form  (see, e.g., Ref.~\cite{Kitajima:2021fpq})
\bae{\label{zetag_prof_mono}
\hat{\zeta}_g(r) = \mu_2\psi_1(r) = \mu_2\frac{\sin(k_* r)}{k_* r}~,
}
and as a result, the non-Gaussian curvature perturbation reads
\bae{
\label{eq: prof_mono}
\hat{\zeta}(r) = -\frac{1}{\gamma}\ln(1 - \gamma \mu_2\frac{\sin(k_* r)}{k_* r}).
} 

\subsection{The Compaction function}
\label{sec:compaction}
In the \ac{FLRW} universe, the spacetime metric for the spherical symmetric perturbation in the long wavelength limit can be written by
\bae{
\dd{s}^2 &= -\dd{t}^2 + a^2(t) \ee^{2\zeta(r)}\left[\dd{r}^2 + r^2 \dd{\Omega}^2\right]=-\dd{t}^2+a^2(t)\ee^{2\zeta(r)}\dd{r}^2+R^2(r)\dd{\Omega}^2,
}
where $a(t)$ is the scale factor and the angular line element is $\dd{\Omega}^2 = \dd{\theta}^2 + \sin^2{\theta}\dd{\phi}^2$ and $R(r)=a\ee^{\zeta(r)}r$ is the areal radius which relates to the proper size of the overdensity region. 
It is known that the compaction function is a useful quantity for estimating whether a peak collapses into a black hole or not~\cite{Shibata:1999zs, Harada:2015yda}.
In the radiation-dominated universe, the compaction function in terms of the curvature perturbation is given by
\bae{\label{eq: compaction}
    \calC(r)=\frac{2}{3}\bqty{1-\qty(1+r\zeta^\prime(r))^2}. 
}
The threshold on the maximal compaction function varies for different profiles of the curvature perturbation. However, it was shown that the threshold on the averaged compaction function~\cite{Escriva:2019phb}
\bae{\label{meanc}
    \bar{\calC}_\um=\left.\pqty{4\pi\int^{R(r_\um)}_0\calC(r)R^2(r)\dd{R(r)}}\middle/\pqty{\frac{4\pi}{3}R^3(r_\um)}\right. ~,
}
is approximately universal on the different shapes of the curvature perturbations and the physical origin of universality is briefly discussed in \cite{Kehagias:2024kgk}.
The radius $r_\um$ corresponds to the extremum of $\calC(r)$ obtained by solving $\zeta^{\prime} + r \zeta^{\prime\prime} = 0$.
According to this prescription, when the mean compaction function $\bar{\calC}_\um$ exceeds the threshold value $\bar{\calC}_\uth=2/5$, the corresponding peak is supposed to collapse into a PBH~\cite{Escriva:2019phb}.

Note that depending on the monotonicity of the areal radius $R(r)$ we define the type I and II fluctuations  (see, e.g., a recent paper~\cite{Uehara:2024yyp}). The monotonicity of $R(r)$
leads to
\bae{
    \dv{R(r)}{r}=a\ee^{\zeta(r)}\left(1+r\zeta^\prime(r)\right)>0,
}
and the initial curvature perturbation $\zeta$ on super-horizon scale is called type I if it satisfies this condition. 
On the other hand, the areal radius $R(r)$ is no longer monotonic for type II fluctuations (i.e., $\dv{R(r)}{r} \propto 1 + r\zeta^{\prime} < 0$). 

\subsection[The $q$-function method]{\boldmath The $q$-function method}
\label{sec:q-func}

It was found that the profile-dependence of the threshold is mainly determined by the second derivative at the maximum of $\mathcal{C}(r)$, which is described by an analytical indicator called ``$q$-function'' in Ref.~\cite{Escriva:2019phb}. In Ref.~\cite{Escriva:2022pnz}, it was found that the $q$-function is also very useful in the non-Gaussian case of quadratic expansion with large negative $f_{\rm NL}$. We will review the $q$-function method in this subsection.

The metric in terms of the comoving areal radius $\tilde{r}(r) = r \ee^{\zeta(r)} = R(r)/a$ can be written as
\bae{
\dd{s}^2 = -\dd{t}^2 + a^2(t)\left\{\frac{\dd{\tilde{r}}^2}{1 - K(\tilde{r}) \tilde{r}^2} + \tilde{r}^2 \dd{\Omega}^2\right\}.
}
The curvature perturbation $\zeta$ and the spacial curvature $K$ are related as
\bae{
\zeta(r) = \int^{\tilde{r}(r)}_{\infty} \frac{\dd{\tilde{r}^{\prime}}}{\tilde{r}^{\prime}}\left(1 -  \frac{1}{\sqrt{1 - K(\tilde{r}^{\prime}) \tilde{r}^{\prime 2}}}\right).
}
The compaction function in the comoving coordinate $r$ can be calculated by Eq.~\eqref{eq: compaction} and the expression in terms of the comoving areal radius $\tilde{r}$ is given by
\bae{
\tilde{\calC}(\tilde{r}) = \calC\qty(r(\tilde{r})) =\frac{2}{3} K(\tilde{r})\tilde{r}^2.
}
If one supposes the following fiducial profile of $K$ characterised by a parameter $q$, as introduced in Ref.~\cite{Escriva:2019phb},
\bae{
\label{eq: fiducial_K}
K_{q}(\tilde{r}) =  \frac{3}{2}\frac{\tilde{\calC}(\tilde{r}_{\rm m})}{\tilde{r}_{\rm m}^2} \exp[\frac{1}{q}\left\{1 - \left(\frac{\tilde{r}}{\tilde{r}_{\rm m}}\right)^{2q} \right\}],
}
where $\tilde{r}_{\rm m} = \tilde{r}(r_{\rm m})$, the characteristic $q$ parameter (or the $q$-function, which is an implicit function of the parameters of the peak profile) is related to the maximal compaction and its second derivative:
\bae{\label{eq: q}
q = -\frac{1}{4}\tilde{r}_{\rm{m}}^2 \frac{\tilde{\calC}^{\prime\prime}(\tilde{r}_{\rm m})}{\tilde{\calC}(\tilde{r}_{\rm m})} = -\frac{1}{4}
r_{\rm{m}}^2
\frac{\calC^{\prime\prime}(r_{\rm m})}{\calC(r_{\rm m})\bqty{1 - (3/2)\calC(r_{\rm m}) }}.
} 
On the other hand, the maximal mean compaction $\bar{\calC}_\um$ can be analytically calculated for the fiducial profile~\eqref{eq: fiducial_K}. It is noticed that $\bar{\calC}_\um$ depends on the equation of state parameter $w$ (see Eq.~(II.35) in Ref.~\cite{Escriva:2022duf} and also Ref.~\cite{Escriva:2020tak}). If one assumes the spacial curvature $K(r)$ given by Eq.~(\ref{eq: fiducial_K}) which is the case of $w = 1/3$, the critical threshold coincides with $\bar{C}_{\rm th} = 2/5$. The criterion $\bar{\calC}_\um=2/5$ can be recast into the threshold on the maximal compaction function $\delta_{\rm c}$ as

\bae{
\delta_{\rm c} = \frac{4}{15}\ee^{-1/q}\frac{q^{1-5/(2q)}}{\Gamma\qty(5/(2q)) - \Gamma(5/(2q), 1/q)},
}
where  $\Gamma(s)$  is the Gamma function and $\Gamma(s, z)$ is the upper incomplete Gamma function.
Interestingly, even for profiles different from the fiducial one \eqref{eq: fiducial_K}, a simple criterion given by 
$\calC(r_{\rm m}) \geq \delta_{\rm c}$
with the $q$-function calculated by Eq.~\eqref{eq: q} somehow works quite well. Ref.~\cite{Escriva:2022pnz} clarified that the corresponding threshold values are consistent with numerical results within 2$\%$ errors in the presence of positive or small negative non-Gaussianities, though the discrepancy with respect to the numerical results becomes larger in the presence of large negative $f_{\rm NL}$ for quadratic-expansion non-Gaussianity. We will show that this holds for logarithmic non-Gaussianity too.

\subsection{Numerical study on the thresholds} \label{sec:numerical_results} 
Though the above two analytical criteria work for quadratic-expansion non-Gaussianity with a small $f_\mathrm{NL}$, they are not ensured to work for the logarithmic non-Gaussianity. In this work, we hence numerically derive the amplitude threshold ($\mu_{2,{\rm c}}$) for \ac{PBH} formation depending on the parameter $\gamma$ in Eq.~(\ref{general_expotail}). For this purpose, we use a numerical relativity code based on the \ac{BSSN} formalism~\cite{Baumgarte_1998, PhysRevD.52.5428} and reformulated for curvilinear coordinates in spherical symmetry as in Ref.~\cite{Alcubierre:2011pkc}. The matter corresponds to a perfect fluid with a barotropic equation of state, $p = \omega \rho$, where $p$ and $\rho$ are the pressure and energy density of the fluid, respectively, and $\omega = 1/3$ for an ultra-relativistic fluid (radiation). The evolution of the matter components is done by solving the general-relativistic hydrodynamic equations as described in Ref.~\cite{10.1093/acprof:oso/9780199205677.001.0001}. See also Ref.~\cite{Staelens:2019sza} for related work on PBH formation.
In contrast to alternative codes based on the Misner--Sharp formalism \cite{Misner:1964je,Polnarev:2006aa,Escriva:2019nsa}, we are able to compute the amplitude threshold for \acp{PBH} forming from type II fluctuations, where Misner--Sharp codes fail because of the appearance of coordinates singularities. In the parameter region where $\gamma \gtrsim -2 $, we have checked and found that both BSSN and Misner--Sharp codes agree, and yield the same results. 

\begin{figure} 
	\centering
	\includegraphics[width=1\hsize]{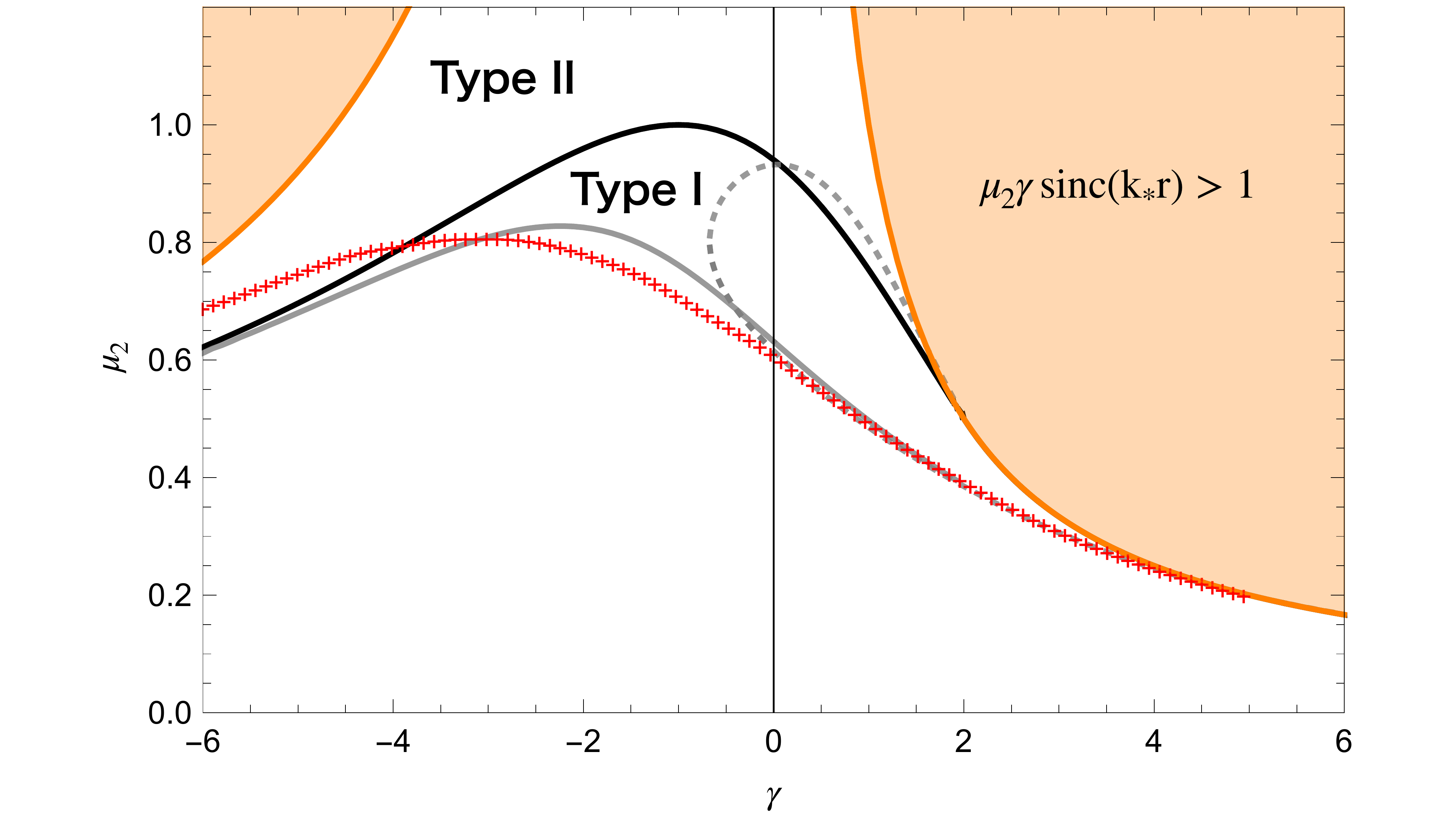}
	\caption{The PBH diagram in terms of the peak amplitude $\mu_2$ in the case of the logarithmic non-Gaussianity. The black solid line denotes the boundary for the type I and type II fluctuations inferred by the monotonicity of the areal radius. 
    The orange lines show the boundary where the inside of the function of Eq.~\eqref{eq: prof_general} coincides with zero. 
    The profile Eq.~\eqref{eq: prof_general} becomes doubtful over the orange lines. The gray curve represents the analytical estimation evaluated by the $q$-function method. The gray dashed curve corresponds to the 
    maximal mean compaction function $\bar{\calC}_\um=2/5$. The red cross points denote the thresholds that form type A \acp{PBH} obtained by the numerical simulation. The type A \ac{PBH} formation from the type I fluctuation happens on the red cross points for $-3.8  
    \lesssim\gamma  
    \lesssim5.0$. The type A \ac{PBH} formation from the type II fluctuation occurs on the red cross points in $ 
    \gamma  
    \lesssim-3.8$.}
	\label{fig: muth_general_exp}
\end{figure}

Our results for the amplitude threshold for PBH formation across different values for $\gamma$ is shown with red crosses in Fig.~\ref{fig: muth_general_exp}. No \ac{PBH} is formed below this curve. The black solid line denotes the boundary for the type I and type II fluctuations computed by the monotonicity of the areal radius. The parameter space below this line corresponds to type I fluctuations, while type II fluctuations span the parameter space above. The gray solid curve represents the analytical estimation of the threshold evaluated by the $q$-function method introduced in Sec.~\ref{sec:q-func}, and the gray dashed curve is obtained from the maximal mean compaction function $\bar{\calC}_\um=2/5$ shown in Sec.~\ref{sec:compaction}. The orange curves correspond to the condition where the argument of the logarithmic function in the peak profile Eq.~\eqref{eq: prof_general} equals zero, 
above which $\zeta$ diverges. This implies the inflaton can not evolve classically to end inflation, as the time duration of inflation (described by $\delta N=\zeta\to\infty$) in these Hubble patches are infinite. Quantum diffusion must be taken into account the study the fate of such \textit{eternal inflation} regions \cite{Vennin:2020kng}, which have complicated global structure and are beyond the scope of this paper. Near the border of the orange region and the Type II region, such long-lived inflating bubbles will be rare and surrounded by classically evolving spacetime, which also collapse into \acp{PBH} finally. Such a bubble channel dominates \ac{PBH} formation when $\gamma\gtrsim3.1$~\cite{Escriva:2023uko}, which we will not study in this paper. Therefore we restrict ourselves in $\gamma\lesssim3.1$. We found that \acp{PBH} formed from both type I and type II fluctuations with amplitudes slightly larger than the threshold are classified as so-called type A, a nomenclature introduced in Ref.~\cite{Uehara:2024yyp}, which corresponds to \ac{PBH} formed through the gravitational collapse of the overdensity region after the curvature perturbation re-enters the Hubble radius. We expect that type B \acp{PBH} introduced in Ref.~\cite{Uehara:2024yyp}, which is classified by the existence of bifurcating trapping horizons (see Fig.~9 in Ref.~\cite{Uehara:2024yyp} and also Ref.~\cite{Shimada:2024eec}), form only at much larger amplitudes.
Then, in terms of the abundance of total \acp{PBH}, we consider
that the contribution from the type B \acp{PBH} would be exponentially suppressed and negligible. Although we do not investigate the details of the formation of the type B \ac{PBH},  interestingly, contrary to what previously thought, we found a region of type II fluctuations where no black hole is formed, that is,  type II fluctuations do not necessarily lead to \ac{PBH} formation (see the region between  red cross points and black solid line for $ \gamma \lesssim -3.8$ in Fig.~\ref{fig: muth_general_exp}).

\subsection{Abundance of \acp{PBH}}
\label{sec: PBH_abundance}

\begin{figure}
    \centering
    \begin{tabular}{c}
        \begin{minipage}[b]{0.5\hsize}
            \centering
            \includegraphics[width=0.95\hsize]{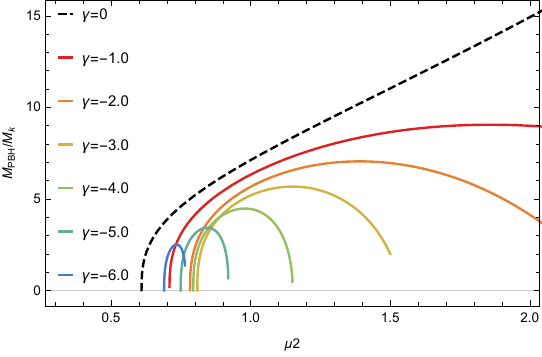}
        \end{minipage}
        \begin{minipage}[b]{0.5\hsize}
            \centering
            \includegraphics[width=0.95\hsize]{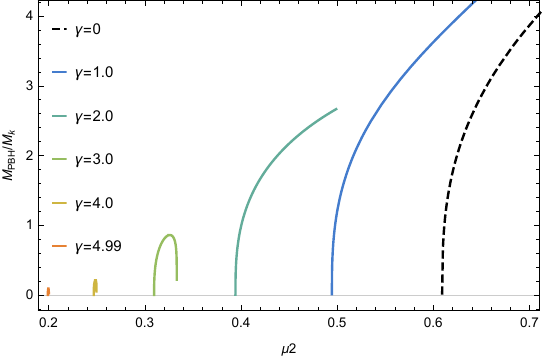}
        \end{minipage}
    \end{tabular}
	\caption{The normalized \ac{PBH} mass as a function of $\mu_2$ for each value of the negative (left) and the positive (right) $\gamma$. Although the \ac{PBH} mass has a local maximum for $\gamma <0$ and $2 < \gamma$, it does not  
    in the range of $0 \leq \gamma \leq 2$.}
	\label{fig: Mpbh}
\end{figure}

\begin{figure}
    \centering
    \begin{tabular}{c}
        \begin{minipage}[b]{0.5\hsize}
            \centering
            \includegraphics[width=0.95\hsize]{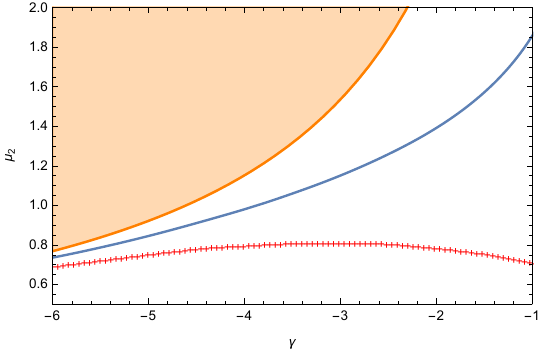}
        \end{minipage}
        \begin{minipage}[b]{0.5\hsize}
            \centering
            \includegraphics[width=0.95\hsize]{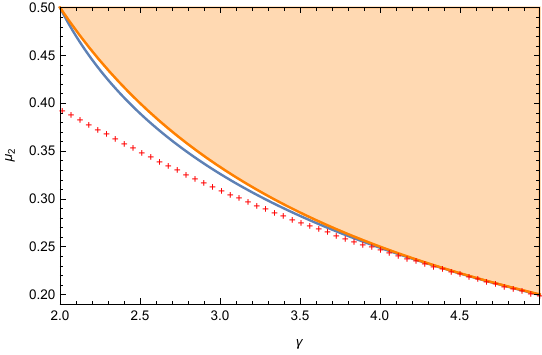}
        \end{minipage}
    \end{tabular}
	\caption{The maximum $\mu_2$ corresponding to the local maximum of \ac{PBH} mass (blue). The orange shaded areas are the doubtful regions,  
    as in Fig.~\ref{fig: muth_general_exp}.  
    Red cross points are the numerical threshold, 
    as in Fig.~\ref{fig: muth_general_exp}.}
	\label{fig: mu_maximize_Mpbh}
\end{figure}

Based on the result of the threshold for 
the \ac{PBH} formation
shown above, in this subsection
we investigate the abundance of \acp{PBH} with logarithmic non-Gaussianity. First of all, we suppose that Type A \acp{PBH} obey critical collapse, such that the resultant mass follows a scaling relation~\cite{Choptuik:1992jv}
\bae{
    M_\PBH=K\qty(\mu_2-\mu_{2,\uc})^p M_H ~,
}
with an universal power $p\simeq0.36$~\cite{Abrahams:1993wa, Evans:1994pj}, an order unity coefficient
$K$, and the threshold  
$\mu_{2,\uc}$. For simplicity, we set $K\simeq1$ in this paper.
$M_H$ is the Hubble mass at the Hubble reentry of the maximal radius, $R(r_\um)H=1$.
Substituting $R(r_{\rm{m}}) = a(t) r_{\rm m}\ee^{\zeta(r_{\rm m})}$, the \ac{PBH} mass can be rewritten as
\bae{\label{eq: Mpbh}
    M_\PBH=K\qty(\mu_2-\mu_{2,\uc})^p(k_* r_{\rm m})^2\ee^{2\zeta(r_{\rm m})} M_{k_*},
}
where $M_{k_*}$ is the Hubble mass at the Hubble reentry of the scale $k_*$ in the background universe given by
\bae{
    M_{k_*} \simeq 10^{20} \left(\frac{g_*}{106.75}\right)^{-1/6}\left(\frac{k_*}{1.56\times 10^{13} \,\si{Mpc^{-1}}}\right)^{-2}~\si{gram}  ~,
}
with the effective degrees of freedom for the energy density $g_*$ at the Hubble crossing. The \ac{PBH} mass
as a function of $\mu_2$
for each $\gamma$ is exhibited in Fig.~\ref{fig: Mpbh}. While the \ac{PBH} mass has a local maximum for $\gamma < 0$ and $2 < \gamma$, it does not in the range of $0 \leq \gamma \leq 2$. 
As we will see later, 
to evaluate the total abundance of \acp{PBH},
we need to perform the integration of the mass function with respect to $\mu_2$ and
set a maximum value of $\mu_2$ for each $\gamma$.  Although we do not know the exact upper bound of $\mu_2$ for the type A \ac{PBH} formation, 
we adopt the value of $\mu_2$ which gives the local maximum of \ac{PBH} mass as the maximum value for $\gamma < 0$ and $2 < \gamma$, which is shown in Fig.~\ref{fig: mu_maximize_Mpbh} as a function of $\gamma$ by the blue lines. On the other hand, we use $1/\gamma$ (corresponding to the orange line of Fig.~\ref{fig: muth_general_exp}) as the maximum value of $\mu_2$ for $0 \leq \gamma \leq 2$.
We have checked that
the total abundance of \acp{PBH} is less sensitive to these choices of the maximum value of $\mu_2$, and this is 
because the \ac{PBH} formation mainly happens near the critical threshold $\mu_{2, \rm c}$.

The peak number density with the amplitude $\mu_2$ can be statistically computed by the \emph{peak theory}~\cite{Bardeen:1985tr}. Using the relation between $\mu_2$ and the PBH mass in Eq.~\eqref{eq: Mpbh}, a peak number density can be recast into the current PBH energy density, which is described by the \ac{PBH} mass function $f_\PBH(M)$, defined as the current \ac{PBH} fraction of the dark matter in the mass range of $[M, M \ee^{\dd{\ln{M}}}]$ as\footnote{It should be noted that the \ac{PBH} abundance expression in the original paper~\cite{Yoo:2018kvb} had a factor $1/27$ error. It has been corrected in our expression.}  
\bae{\label{eq: fpbh}
    &f_\PBH(M)=\frac{\rho_{\PBH,0}}{\rho_{{\rm DM},0}} \\* \nonumber
    &=\left(\frac{\Omega_{\DM,0} h^2}{0.12}\right)^{-1}\left(\frac{M}{10^{20} \,\si{g}}\right)\left(\frac{k_*}{1.56\times 10^{13} \,\si{Mpc^{-1}}}\right)^{3} \left(\frac{\abs{\frac{\dd{\ln{M}}}{\dd{\mu_2}}}^{-1}f\left(\frac{\mu_2(M)}{\sqrt{A_g}}\right)\calN(\mu_2(M), \sqrt{A_g})}{1.4\times 10^{-14}}\right),
}
where the subscript ``0'' denotes the value at present, $\Omega_{\DM}$ is the density parameter of the dark matter,  $\calN(x, \sigma) = \frac{1}{\sqrt{2\pi}\sigma} \exp\left(-\frac{x^2}{2 \sigma^2}\right)$ denotes the zero-mean Gaussian distribution with the variance $\sigma$, and the function $f(\xi)$  corresponds to  
\bme{
    f(\xi)=
    \frac{1}{2}\xi(\xi^2-3)\pqty{\erf\bqty{\frac{1}{2}\sqrt{\frac{5}{2}}\xi}+\erf\bqty{\sqrt{\frac{5}{2}}\xi}} \\
    +\sqrt{\frac{2}{5\pi}}\Bqty{\pqty{\frac{8}{5}+\frac{31}{4}\xi^2}\exp\bqty{-\frac{5}{8}\xi^2}+\pqty{-\frac{8}{5}+\frac{1}{2}\xi^2}\exp\bqty{-\frac{5}{2}\xi^2}}.
}
The total \ac{PBH} abundance is obtained by integrating Eq.~(\ref{eq: fpbh})
\bae{\label{eq: fPBHtot}
    f_\PBH^\tot=\int f_\PBH(M)\dd{\ln M}.
}

\begin{figure} 
    \centering
    \begin{tabular}{c}
        \begin{minipage}[b]{0.5\hsize}
            \centering
            \includegraphics[width=0.95\hsize]{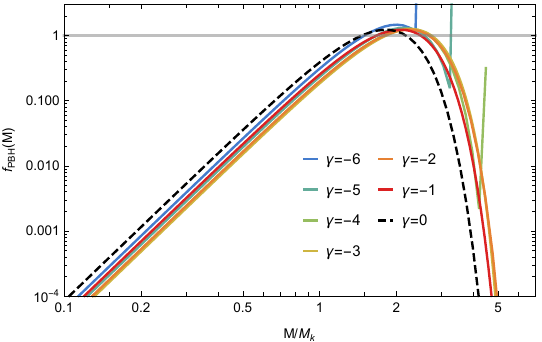}
        \end{minipage}
        \begin{minipage}[b]{0.5\hsize}
            \centering
            \includegraphics[width=0.95\hsize]{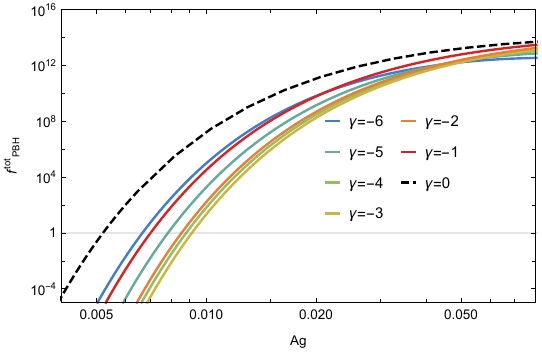}
        \end{minipage}
    \end{tabular}
	\caption{\emph{Left}: the \ac{PBH} mass spectra for 
    $(A_g, \gamma) = (6.69 \times 10^{-3}, -6)$, $(7.86 \times 10^{-3}, -5)$, $(8.81\times 10^{-3},-4)$, $(9.14\times 10^{-3},-3)$, $(8.56\times 10^{-3},-2)$, $(7.04\times 10^{-3},-1)$, and $(5.21\times 10^{-3},0)$ from blue to red and the black dashed line.
    The amplitude $A_g$ is chosen for each $\gamma$ so that the total \ac{PBH} abundance $f_\PBH^\tot$ becomes unity.
    \emph{Right}: the total \ac{PBH} abundance~\eqref{eq: fPBHtot} for each $\gamma$ with the same color code as the left panel.
    }
	\label{fig: fpbh_fpbhtot_Gumbel}
\end{figure}

\begin{figure} 
    \centering
    \begin{tabular}{c}
        \begin{minipage}[b]{0.5\hsize}
            \centering
            \includegraphics[width=0.95\hsize]{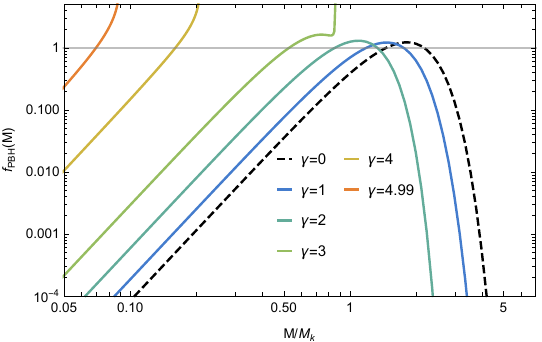}
        \end{minipage}
        \begin{minipage}[b]{0.5\hsize}
            \centering
            \includegraphics[width=0.95\hsize]{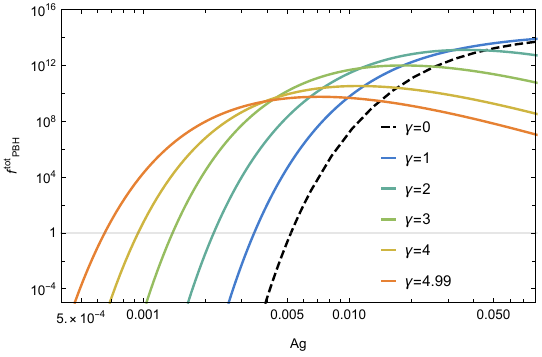}
        \end{minipage}
    \end{tabular}
	\caption{\emph{Left}: the \ac{PBH} mass spectra for 
    $(A_g, \gamma) = (5.21 \times 10^{-3}, 0)$, $(3.46 \times 10^{-3}, 1)$, $(2.20\times 10^{-3},2)$, $(1.39\times 10^{-3},3)$, $(9.45\times 10^{-4},4)$, and $(6.52\times 10^{-4},4.99)$ by the black dashed line and from blue to yellow.
    They are again normalized by $f_\PBH^\tot=1$.
    \emph{Right}: the total PBH abundance 
    for the same $\gamma$ values as the left panel.}
	\label{fig: fpbh_fpbhtot_exp}
\end{figure}

The mass spectrum and the total PBH abundance for the negative and the positive $\gamma$ are shown in Fig.~\ref{fig: fpbh_fpbhtot_Gumbel} and Fig.~\ref{fig: fpbh_fpbhtot_exp} respectively. Though  
the \ac{PBH} mass spectrum  
shows a divergent feature for $\gamma < 0$ 
and $2 < \gamma$ at the local maximum of the \ac{PBH} mass due to the Jacobian $\abs{\dv{\ln{M}}{\mu_2}}^{-1}$, the total \ac{PBH} abundance as its integral is convergent thanks to the conservation of the probability. Here, we neglected the contribution from the \ac{PBH} formation in vacuum bubbles but it should be taken into account numerically for a more precise estimation of the \ac{PBH} mass spectrum, especially for $\gamma  
\gtrsim3.1$~\cite{Escriva:2023uko}. In this work, we focus solely on type A PBH formation and leave this point for future work.

\begin{figure}
	\centering
	\includegraphics{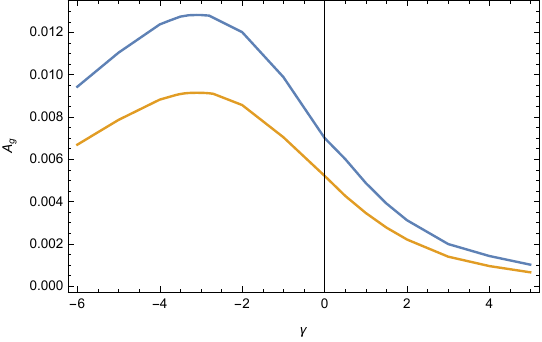}
	\caption{The $\gamma$ dependence of the required amplitude of the Gaussian fluctuations $A_g$ to realize $f_{\rm PBH}^{\rm tot} = 1$ for each fixed \ac{PBH} mass scale $M_{k_*} = 10^{20}\,\si{g}$ (blue) and $\SI{5.45e28}{g}$ (yellow). These mass scales correspond to frequencies $f_* = \SI{0.023}{Hz}$ (blue) and $10^{-6}\,\si{Hz}$ (yellow) respectively.}
	\label{fig: gamma_Ag}
\end{figure}

In Fig.~\ref{fig: gamma_Ag}, we show the $\gamma$ dependence of the required amplitude of the Gaussian fluctuations $A_g$ to realize $f^{\rm tot}_{\rm PBH} =1$ for each fixed \ac{PBH} mass scale $M_{k_*} = 10^{20}\,\si{g}$ (blue), and $\SI{5.45e28}{g}$ (yellow).
As shown in this figure, the required value of $A_g$ decreases  for the positive $\gamma$ since the positive non-Gaussianities promote \ac{PBH} productions. On the other hand, towards the negative semi-axis of $\gamma$, the required $A_g$ increases at first until it reaches its maximum at $\gamma \simeq -3.1$, and then decreases.  
While a negative non-Gaussianity suppresses the \ac{PBH} formation between $0 > \gamma \gtrsim -3.1$, somehow counterintuitively, we find that \ac{PBH} formation again becomes enhanced for larger negative non-Gaussianities below $\gamma \lesssim -3.1$.  This trend is because the critical threshold $\mu_{2,\rm c}$  
increases as $\gamma$ decreases, reaches the maximum at $\gamma \simeq -3.1$, and again decreases which is confirmed down to $\gamma = -6.0$. 
Note that, as we will see later, the mass scales
chosen in Fig.~\ref{fig: gamma_Ag} correspond to typical frequency of \ac{LISA} $f_* = \SI{0.023}{Hz}$ and the peak frequency $f_* = 10^{-6}\,\si{Hz}$ that is compatible with \ac{NANOGrav} observation when one tries to fit the  \ac{NANOGrav} data with \ac{SIGW}, respectively. 
In principle, the value of $\gamma$ which maximizes $A_g$ depends slightly on the peak frequency which in turn is determined by the mass scale and number of relativistic degrees of freedom at the horizon reentry. 
We confirmed that its dependence was tiny and takes maximum value at $\gamma \simeq -3.1$ almost irrespective of the peak frequency.

\section{The scalar induced gravitational wave signals}
\label{sec: SIGW}

\subsection{Formulation}

We will now briefly review the \acp{SIGW} in the presence of primordial non-Gaussianities based on Refs.~\cite{Espinosa:2018eve, Kohri:2018awv, Pi:2020otn, Adshead:2021hnm, Abe:2022xur}.
The perturbed metric including the gravitational waves can be written as
\bae{
    \dd{s}^2 = a(\tau)^2\left[-\qty(1+2\Phi(\tau,\bfx))\dd{\tau}^2 + \left(\qty(1-2\Phi(\tau,\bfx))\delta_{ij}+\frac{1}{2}h_{ij}(\tau,\bfx)\right)\dd{x}^i\dd{x}^j \right],
}
where $\tau$ represents the conformal time, $\Phi(\tau,\bfx)$ is the curvature perturbation in the Newtonian gauge, and $h_{ij}(\tau,\bfx)$ is the transverse traceless tensor perturbation. 
The tensor perturbation can be expanded by the Fourier modes $\hk{}{}$ as
\bae{
     h_{ij}(\tau,\bfx)=\sum_{\lambda = +,\times}\int\dk \ee^{i\bfk\cdot\bfx}e^{\lambda}_{ij}(\bfk)\hk{}{},
     \label{eq:fourier_strain}
}
where $e^{\lambda}_{ij}(\bfk)$ denotes the polarization tensors which satisfy the transverse traceless conditions $e^{\lambda}_{ij}
e^{\lambda'}_{ij}= \delta^{\lambda\lambda^\prime}$  
and $e^{\lambda
}_{ii}=0$.
The tensor power spectrum $P_{\lambda \lambda^{\prime}}(\tau,k)$ is defined as
\bae{
    \braket{\hk{}{}\hk{\prime}{}}=(2\pi)^3 \delta^3(\bfk+\bfk^{\prime})P_{\lambda \lambda^{\prime}}(\tau,k),
}
and the dimensionless power spectrum $\PP_{\lambda\lambda^{\prime}}(\tau,k)$ 
is given by
\bae{
   \PP_{\lambda\lambda^{\prime}}(\tau,k) = \frac{k^3}{2\pi^2}P_{\lambda \lambda^{\prime}}(\tau,k).
}
The density parameter per logarithmic wavenumber is given by 
\bae{
    \Omega_{\GW}=\frac{1}{\rho_{\mathrm{crit}}} 
    \dv{\rho_\GW}{\log k},
}
where $\rho_{\mathrm{crit}}$ and $\rho_{\GW}$ are the critical energy density and the energy density of \acp{GW} respectively. The energy density parameter of the \acp{SIGW} is given by
\bae{\label{eq: OmegaGW}
\Omega_{\GW}(\tau, k) =\frac{1}{48}\lr{\frac{k}{\mathcal{H}(\tau)}}^2 
\sum_{\lambda,\lambda' = +,\times}\overline{\PP_{\lambda\lambda'}(\tau,k)}\,,
}
where $\calH=\partial_\tau\ln a$ is the conformal Hubble parameter and the overline means the oscillation average.  We denote a constant \ac{GW} density parameter in a deep subhorizon limit during the \ac{RD} era as $\Omega_\GW^\RD(k)$, and then the present density parameter can be described as~\cite{Ando:2018qdb}
\bae{
\label{eq: Omg_tot}
\Omega_{\GW} (\tau_0, k)  = \Omega_{\ur,0}\frac{g^*}{g^*_{0}} \left(\frac{g^*_{s,0}}{g^*_{s}} \right)^{4/3}
\Omega_\GW^\RD (k)\,,
}
where $\Omega_\ur$ is the density parameter of the radiation component, and $g^*$ and $g^*_s$ respectively represent the effective number of relativistic degrees of freedom for the energy and entropy densities. We adopt the standard model values $g^* = g^*_s = 106.75, ~ g_0^*=3.384$, and $g_{s,0}^*=3.938$.

Up to the first order in tensor and the second order in scalar, the \ac{EoM} for the Fourier mode of \acp{GW} is given by
\bae{
    \hk{}{\prime\prime} + 2\HH(\tau) \hk{}{\prime} +k^2\hk{}{}=4S_\lambda(\tau,\bfk)\,,
    \label{eq:h}
}
where $S_\lambda(\tau,\bfk)$ is the source term quadratic in $\Phi$, and the prime denotes the partial derivative with respect to the conformal time. The source term can be written as
\bae{\label{eq:source}
    S_\lambda(\tau,\bfk)=\iq{}\Q{}{}f(\abs{\bfk-\bfq},q,\tau)\zeta(\bfq)\zeta(\bfk-\bfq),
}
in terms of the comoving curvature perturbation $\zeta(\bfk)$
with the projection function $\Q{}{}$ which reads
\bae{
    Q_\lambda(\bfk,\bfq)=e^\lambda_{ij}(\bfk)q^iq^j=\frac{q^2}{\sqrt{2}}\sin^2\theta\times\bce{
        \cos(2\psi) & (\lambda=+), \\
        \sin(2\psi) & (\lambda=\times),
    }
}
in the spherical coordinate where the \ac{GW}'s propagation direction $\bfk$ is set as the $z$-direction. The source factor $f(p,q,\tau)$ in the \ac{RD} universe is given by
\bae{
    f(p,q,\tau)&=3\Phi(p\tau)\Phi(q\tau)+ \dv{\Phi(p\tau)}{\ln (p\tau)} \dv{\Phi(q\tau)}{\ln (q\tau)} +\pqty{\Phi(p\tau)\dv{\Phi(q\tau)}{\ln(q\tau)}+\dv{\Phi(p\tau)}{\ln (p\tau)}\Phi(q\tau)}, 
}
where $\Phi (x)$ with a single argument is the linear transfer function which relates the primordial curvature perturbation $\zeta(\bfk)$ and the Newton potential $\Phi(\tau,\bfk)$ by
\bae{\label{eq: relation_phi_zeta}
    \Phi(\tau,\bfk)=\Phi(k\tau)\zeta(\bfk)\,,
}
with
\bae{
    \Phi(x)=-\frac{2}{3}\frac{9}{x^2}\pqty{\frac{\sin(x/\sqrt{3})}{x/\sqrt{3}}-\cos(x/\sqrt{3})}\,,
}
in the \ac{RD} era. The solution of the \ac{EoM} can be constructed by the Green function as 
\bae{\label{eq:hk}
    \hk{}{}=\frac{4}{a(\tau)}\int^\tau\dd{\tilde{\tau}}G_\bfk(\tau,\tilde{\tau})a(\tilde{\tau})S_\lambda(\tilde{\tau},\bfk),
}
where $ G_\bfk(\tau,\tilde{\tau})=\frac{\sin k(\tau-\tilde{\tau})}{k}$ is the tensor Green function.
Combining Eq.~\eqref{eq:source} and Eq.~\eqref{eq:hk}, one finds the two-point function of \acp{SIGW} as
\bme{\label{eq:hh}
    \braket{h_{\lambda_1}(\tau,\bfk_1)h_{\lambda_2}(\tau,\bfk_2)}=\iq{1}\iq{2}\Q{1}{1}\Q{2}{2} \\
    \times I_k(\abs{\bfk_1-\bfq_1},q_1,\tau)I_k(\abs{\bfk_2-\bfq_2},q_2,\tau) 
    \braket{\zeta(\bfq_1)\zeta(\bfk_1-\bfq_1)\zeta(\bfq_2)\zeta(\bfk_2-\bfq_2)},
}
with the kernel function
\bae{
    I_k(p,q,\tau) = 4\int^\tau\dd{\tilde{\tau}}G_\bfk(\tau,\tilde{\tau})\frac{a(\tilde{\tau})}{a(\tau)}f(p,q,\tilde{\tau}).
}

Based on the previous works~\cite{Adshead:2021hnm, Abe:2022xur}, by introducing the perturbative description of the local type non-Gaussianity as
\bae{\label{eq: local_nonG}
\zeta(\bfx)= \zeta_g(\bfx) + F_{\rm NL} \zeta_g^2(\bfx) +   G_{\rm NL} \zeta_g^3(\bfx) + H_{\rm NL} \zeta_g^4(\bfx)+ I_{\rm NL} \zeta_g^5(\bfx) + \cdots, 
}
with the non-linearity parameters $F_{\rm NL}$, $G_{\rm NL}$, $H_{\rm NL}$, 
$I_{\rm NL}$, etc.,
we can perturbatively evaluate the \acp{SIGW} in the presence of the primordial non-Gaussianity.
In the case of the logarithmic non-Gaussianity parameterized by $\gamma$, the perturbative non-linear parameters are respectively given as  
$F_{\rm NL} = 
\gamma/2$, $G_{\rm NL} = 
\gamma^2/3$, $H_{\rm NL}= 
\gamma^3/4$,  
$I_{\rm NL}=
\gamma^4/5$, and so on.  Substituting Eq.~(\ref{eq: local_nonG}) into Eq.~\eqref{eq:hh} and assuming the monochromatic power spectrum for the Gaussian fluctuations~\eqref{eq: monochromatic P}, one can obtain non-Gaussian corrections against the energy density of the \acp{GW} order by order of $A_g$. The total amplitude of the energy density parameter $\Omega_{\rm GW}$ is determined by the combination of the non-linearity parameters and the amplitude of the Gaussian fluctuations, for example, it contains terms proportional to $A_g^2$, $F_{\rm NL}^2 A_g^3$, $(F_{\rm NL} A_g)^4$, and so on. 
For details of the calculations, see Ref.~\cite{Abe:2022xur}.

\subsection{\ac{SIGW} associated with \ac{PBH} dark matter}

\begin{figure} 
	\centering
	\includegraphics{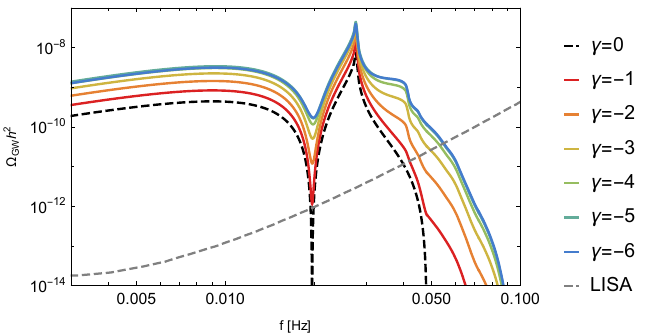}
	\caption{The \ac{SIGW} spectrum for each $\gamma$ value up to $\calO(A_g^4)$ as in Ref.~\cite{Abe:2022xur}. We assume 
    a Dirac-delta power spectrum for the Gaussian fluctuations on $k_*/\si{Mpc^{-1}} = 1.56 \times 10^{13}$ corresponding to $M_{k_*}=10^{20}\,\si{g}$. The gray dashed line represents the LISA sensitivity curve assuming one-year observation~\cite{Schmitz:2020syl}. We used reference values 
    $(A_g, \gamma) = (6.69 \times 10^{-3}, -6)$, $(7.86 \times 10^{-3}, -5)$, $(8.81\times 10^{-3},-4)$, $(9.14\times 10^{-3},-3)$, $(8.56\times 10^{-3},-2)$, $(7.04\times 10^{-3},-1)$, and $(5.21\times 10^{-3},0)$ for the plots so that $f_\PBH^\tot=1$.}
	\label{fig: GW_Gumbel}
\end{figure}

\begin{figure} 
	\centering
	\includegraphics{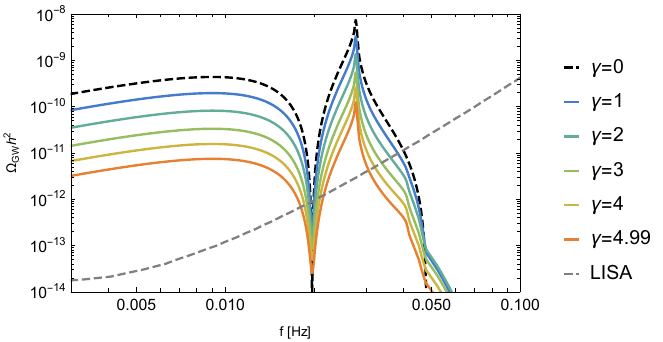}
	\caption{
    The same plot as Fig.~\ref{fig: GW_Gumbel} but for positive $\gamma$'s.
    We used reference values $(A_g, \gamma) = (5.21 \times 10^{-3}, 0)$, $(3.46 \times 10^{-3}, 1)$, $(2.20\times 10^{-3},2)$, $(1.39\times 10^{-3},3)$,  $(9.45\times 10^{-4},4)$, and $(6.52\times 10^{-4},4.99)$. 
    }
	\label{fig: GW_exp}
\end{figure}

The resultant \ac{SIGW} signals for the logarithmic non-Gaussianity are shown in Fig.~\ref{fig: GW_Gumbel} and Fig.~\ref{fig: GW_exp}, supposing monochromatic Gaussian fluctuations which have a peak at $k_*=\SI{1.56e13}{Mpc^{-1}}$ corresponding to $M_{k_*}=10^{20}\,\si{g}$ compatible with the \ac{PBH} \ac{DM} scenario. 
The signals are compared with the \ac{LISA} sensitivity. For the positive $\gamma$ shown in Fig.~\ref{fig: GW_exp}, the required \ac{SIGW} amplitude to realize $f_{\rm PBH}^\tot =1$ is suppressed as $\gamma$ becomes larger since the positive non-Gaussianity promotes \ac{PBH} productions, requiring less typical amplitude of the curvature perturbations. As in Appendix~\ref{Appendix: convergence_SIGW}, the spectrum of \ac{SIGW} is well converged and higher order corrections are subdominant for the positive $\gamma$.
Although the amplitude is attenuated, the \ac{SIGW} signals compatible with the \ac{PBH} \ac{DM} are large enough to be detectable in the \ac{LISA} sensitivity. 
For the negative $\gamma$ in Fig.~\ref{fig: GW_Gumbel}, 
the \ac{SIGW} signals in the \ac{PBH} \ac{DM} scenario increases as $\gamma$ is smaller for $-3 
\lesssim\gamma < 0$, and they would be detectable in \ac{LISA}.
This behavior is similar to that of the required amplitude of the Gaussian fluctuations $A_g$ shown in Fig.~\ref{fig: gamma_Ag}.
For the case with $ \gamma  
\lesssim-3$, 
the \ac{SIGW} signals appear to be saturated for the changes in $\gamma$ around $-5  
\lesssim\gamma  
\lesssim-4$.
However, we found that
we have to be concerned about the validity of the perturbative calculation for the \ac{SIGW}.
As discussed in Appendix~\ref{Appendix: convergence_SIGW},
while the higher-order collections in the \ac{SIGW} spectra are well converged for $-4 \lesssim \gamma \leq 0$, the perturbative calculation has failed due to relatively large non-Gaussianity for $\gamma 
\lesssim-4$.
In this regime, some non-perturbative schemes should be required to accurately compute the \ac{SIGW}, though they are beyond the scope of this paper.

\subsection{\ac{SIGW} in light of recent PTA results}

\begin{figure} 
	\centering
     \includegraphics[width=0.8\hsize]{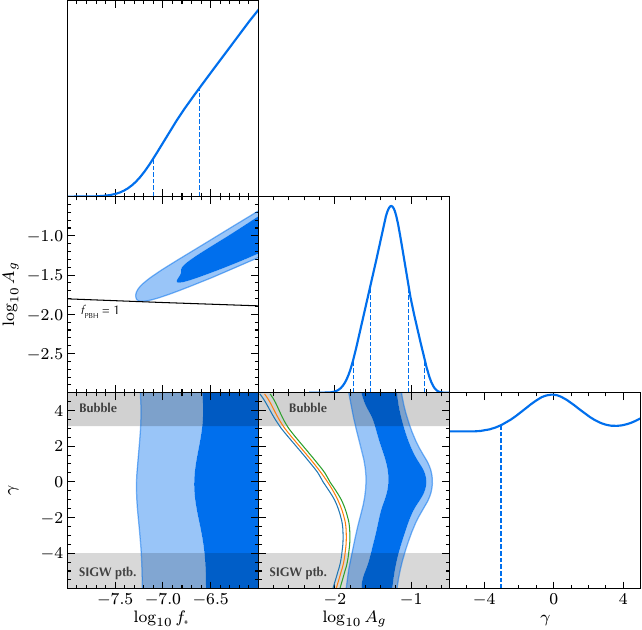}
	\caption{The posterior plot generated by the NANOGrav result. For \acp{SIGW}, we include contributions up to $\calO(A_g^3)$. The black solid line in the $A_g$-$f_*$ figure denotes $f_{\rm PBH}^\tot=1$ for $\gamma = -3.1$, above which the \acp{PBH} are overproduced. The color curved in $\gamma$-$A_g$ figure also represent $f_{\rm PBH}^\tot = 1$ with fixed $f_*/\si{Hz} = 10^{-8}$ (green), $10^{-7}$ (orange), and $10^{-6}$ (blue). The gray-shaded regions represent doubtful regimes 
    where the \ac{PBH} formations from vacuum bubbles are dominated ($\gamma\gtrsim3.1$) or the perturbative computation of \acp{SIGW} fails ($\gamma\lesssim-4$). We used the public code \textsf{PTArcade}~\cite{Mitridate:2023oar} for the plot.}
	\label{fig: NANOGrav_param_estim}
\end{figure}

We also discuss the compatibility of the recent reports about the nHz \ac{SGWB} by the \ac{PTA} experiments (the NANOGrav 15-year data in particular)~\cite{NANOGrav:2023hvm, EPTA:2023fyk, Reardon:2023gzh, Xu:2023wog} with \acp{SIGW} in light of the logarithmic non-Gaussianity.
We have performed the parameter estimation against our model with the public Python module \textsf{PTArcade}~\cite{Mitridate:2023oar}. We consider contributions up to $A_g^3$ and include the term proportional to $G_{\rm NL}$ which was not contained in the previous analysis~\cite{Franciolini:2023pbf}.  Our prior choice is displayed in Table~\ref{table: prior} and the resultant posterior plot is shown in Fig.~\ref{fig: NANOGrav_param_estim}.
The color curves represent the $\gamma$ dependence of the required $A_g$ to realize $f_{\rm PBH}^{\tot} =1$ for each reference peak frequency $f_*/\rm Hz = 10^{-8}$~(green), $10^{-7}$~(orange), and $10^{-6}$~(blue). 
The \ac{PBH} overproduction happens if $A_g$ is larger than these lines. We used the scale and the peak frequency relation which is given by 
\bae{
    f_* = 10^{-9}\left(\frac{k_*}{\SI{6.68e5}{Mpc^{-1}}}\right) \,\si{Hz}.
}
The gray-shaded regions represent the limitations of our analysis. In the upper shaded region ($\gamma\gtrsim3.1$), the PBHs formed in the bubble channel become dominant \cite{Escriva:2023uko}, which are completely neglected in our calculation. Carefully taking such PBHs into account will drive all the colored curves in this region to the left, and make it even more difficult to fit the data. On the other hand, in the lower shaded region ($\gamma\lesssim-4$), the high-order terms contributes more to the \ac{SIGW}, thus our calculation up to $\mathcal{O}(A_g^3)$ underestimates $\Omega_\mathrm{GW}$. The posterior $A_g$ contour after considering all higher-order contributions might move to the left, yet we do not have a satisfying method to calculate them systematically. Nevertheless, such a trend implies that it might be possible to interpret nHz \ac{SGWB} as the \ac{SIGW} in the logarithmic non-Gaussianity with large and negative $\gamma\lesssim-4$. However, one of the main results of our paper is that such an interpretation inevitably have the PBH overproduction problem when $\gamma\gtrsim-4$. 
This is clearly shown by the black solid line in Fig.~\ref{fig: NANOGrav_param_estim}, which corresponds to $f_{\rm PBH}^\tot =1$ for $\gamma = -3.1$ where the required amplitude $A_g$ becomes maximum for the changes in $\gamma$. \acp{PBH} are overproduced even in this case, as the 2-$\sigma$ contour of the posterior $A_g$ is still above this line, and such tension is more serious for other values of $\gamma$, as long as $\gamma\gtrsim-4$.  
Note that our result supposes that all observed \ac{SGWB} are caused by \acp{SIGW} and the result can be  
milder if one allows other possible sources such as the super-massive black hole mergers (e.g., see Fig.~7 in Ref.~\cite{NANOGrav:2023hvm}). Also, we assume a monochromatic power spectrum and a radiation-dominated background, while the tension can be alleviated by either a broad power spectrum or by a stiff equation-of-state, both of which can suppress the PBH formation even in the Gaussian case \cite{Iovino:2024tyg, Domenech:2024rks}.

\begin{table} 
    \centering
    \begin{tabular}{ll}
        \hline\hline
        Parameters & Prior\\
        \hline
        $\log_{10}A_g$ & $\text{LogUniform}~[-4, ~1]$\\
        $\log_{10}f_* /[\rm Hz]$  &  $\text{LogUniform}~[-8, ~-6]$\\
        $\gamma$ & $\text{Uniform}~[-6.0, ~4.99]$\\
        \hline\hline
    \end{tabular}
    \caption{Prior distributions used for the parameter estimation. $f_*$ is the peak frequency of the primordial scalar power spectrum which is chosen within the suitable frequency range for NANOGrav observation.}
    \label{table: prior}
\end{table}

\section{Conclusions}
\label{sec: conclusion}
In this paper, we have investigated the \ac{PBH} formation and the associated \ac{SIGW} signals with the logarithmic non-Gaussianity which is motivated in a wide class of inflation models. By using the numerical relativistic simulation, we first investigated the threshold for the type A \ac{PBH} formation with respect to the peak height $\mu_2$ and the parameter $\gamma$ which characterizes the non-Gaussian property of the primordial perturbation. Our results for the threshold 
for \ac{PBH} formation across different values for $\gamma$ are shown with red cross points in Fig.~\ref{fig: muth_general_exp}. 
We have found that PBHs formed from both type I and type II fluctuations with amplitudes
slightly larger than the threshold are classified as so-called type A, a nomenclature introduced in Ref.~\cite{Uehara:2024yyp}.
We also found a region where no \ac{PBH} is formed from type II fluctuations contrary to the previous thought and concluded that type II fluctuation is not necessarily accompanied by \ac{PBH} formation. 

By using the resultant threshold depending on the non-linearity parameter $\gamma$, we have calculated the \ac{PBH} mass spectrum and the total \ac{PBH} abundance. Our results for the negative and the positive $\gamma$ are shown in Fig.~\ref{fig: fpbh_fpbhtot_Gumbel} and Fig.~\ref{fig: fpbh_fpbhtot_exp} respectively. Though 
the \ac{PBH} mass spectrum shows a divergent feature for $\gamma < 0$ 
and $2 < \gamma$ at the local maximum of the \ac{PBH} mass due to the Jacobian $\abs{\dv{\ln{M}}{\mu_2}}^{-1}$, the total \ac{PBH} abundance as its integral is convergent thanks to the conservation of the probability. Although we neglected the contribution from the \ac{PBH} formation in vacuum bubbles, it is needed to take into account numerically for a more precise estimation of the \ac{PBH} mass spectrum in $\gamma \gtrsim3.1$~\cite{Escriva:2023uko}. In this work, we focus solely on type A \ac{PBH} formation and leave this point for future work. 
The required amplitude $A_g$ to realize $f^{\rm tot}_{\rm PBH} =1$,
that is, in \ac{PBH} \ac{DM} scenario, for each fixed \ac{PBH} mass scale is shown in Fig.~\ref{fig: gamma_Ag}. The value of $A_g$ decreases for the positive $\gamma$ when the total \ac{PBH} abundance is fixed in unity since the positive non-Gaussianities promote \ac{PBH} productions.
On the other hand, $A_g$ value increases for the negative $\gamma$, takes maximum at $\gamma \simeq -3.1$, and decreases again at least down to $\gamma = -6.0$. While the negative non-Gaussaianities suppress the \ac{PBH} formations for $0 > \gamma 
\gtrsim -3.1$, counterintuitively, the enhancement of \ac{PBH} formations happens for $\gamma \lesssim-3.1$.  This feature comes from the similar tendency of the threshold as shown in Fig.~\ref{fig: muth_general_exp}.

We have also computed the \ac{SIGW} signals in \ac{PBH} \ac{DM} scenario, that is, $f_{\rm PBH}^\tot = 1$. 
The resultant \ac{SIGW} signals with the logarithmic non-Gaussianity are shown in Fig.~\ref{fig: GW_Gumbel} and Fig.~\ref{fig: GW_exp}, supposing monochromatic Gaussian curvature perturbations on $k_*=\SI{1.56e13}{Mpc^{-1}}$ corresponding to $M_{k_*}=10^{20}\,\si{g}$ compatible with the \ac{PBH} \ac{DM} scenario. 
In the positive $\gamma$ region, the \ac{SIGW} amplitude corresponding to $f_{\rm PBH}^\tot =1$ is suppressed as $\gamma$ becomes larger since the positive non-Gaussianity 
promotes \ac{PBH} productions, requiring less scalar perturbations.
In the calculation of \acp{SIGW}, we employ the perturbative approach discussed in Ref.~\cite{Abe:2022xur}, confirming that the \ac{SIGW} spectra are well converged and higher order corrections are subdominant for the positive $\gamma$.  
Although the amplitude is attenuated, the \ac{SIGW} signals are large enough to be detectable in the LISA sensitivity. 
For negative $\gamma$ case, the required scalar perturbation is larger, which predicts higher \ac{SIGW} than the Gaussian case, easier to be detected by LISA.  
However, we found that we have to be concerned about the validity of the perturbative calculation for the \ac{SIGW}. While the higher-order contributions to the \ac{SIGW} spectra are well converged for $-4 \lesssim \gamma \leq 0$, the perturbative calculation has failed due to the existence of relatively large non-Gaussianity for $\gamma 
\lesssim-4$. 
Non-perturbative schemes are required to accurately compute the \ac{SIGW} in this regime.

Finally, we discussed the possibility of explaining the \ac{SGWB} signal reported by the recent \ac{PTA} experiments by the \acp{SIGW} with the logarithmic non-Gaussianity. The resultant posterior plot for the model parameters is exhibited in Fig.~\ref{fig: NANOGrav_param_estim}. Although there are suitable parameter regions consistent with the NANOGrav data, \ac{PBH} overproduction is a serious problem. However, it is implied that higher-order contributions when $\gamma\lesssim-4$ might alleviate the tension, yet the non-perturbative schemes to calculate the \ac{SIGW} is necessary to study this problem quantitatively.  

\vspace{1em}
\noindent
\textbf{Note added:} At the stage where our results are being compiled into this paper, we noticed that Masaaki Shimada, Albert Escriv\`{a}, Daiki Saito, Koichiro Uehara, and Chul-Moon Yoo were working on similar issues \cite{Shimada:2024eec}. The works of our groups were conducted independently, and we agreed to submit our papers to arXiv on the same day.

\acknowledgments

We thank the authors of Ref.~\cite{Shimada:2024eec} for discussions on the type B \ac{PBH} formation scenario. We also thank Jaume Garriga and Richard Bond for useful comments. This work is supported in part by the National Key Research and Development Program of China Grant No.~2020YFC2201502 (SP),
and by JSPS KAKENHI Grants
No.~JP22K03639 (HM), JP24K00624 (SP), JP24K07047 (YT), JP20K03968 (SY), JP24K00627 (SY), and JP23H00108 (SY). 
RI is supported by JST SPRING, Grant No.~JPMJSP2125, and the ``Interdisciplinary Frontier Next-Generation Researcher Program of the Tokai Higher Education and Research System''.
CJ is funded by the National Natural Science Foundation of China (NSFC) grant No.~12347132, and NSFC RFIS grant No.~W2433007. 
SP is supported by NSFC grant No.~12475066 and No.~12047503. 
SP and SY are also supported by the World Premier International Research Center Initiative (WPI Initiative), MEXT, Japan.

\appendix

\section{The convergence of the scalar-induced gravitational waves}
\label{Appendix: convergence_SIGW}

\begin{figure} 
    \centering
    \begin{tabular}{c}
        \begin{minipage}[b]{0.5\hsize}
            \centering
            \includegraphics[width=0.95\hsize]{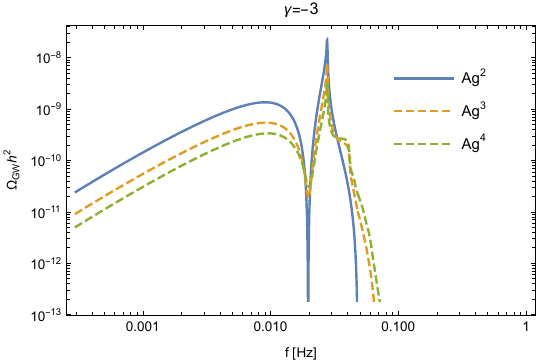}
        \end{minipage}
        \begin{minipage}[b]{0.5\hsize}
            \centering
            \includegraphics[width=0.95\hsize]{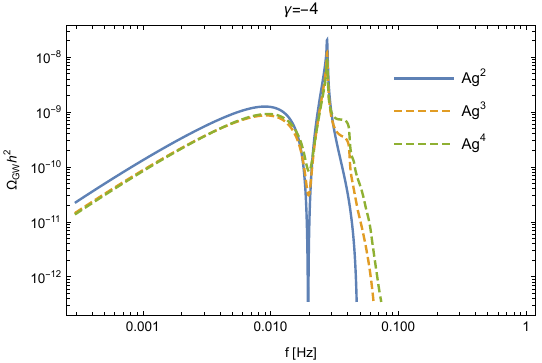}
        \end{minipage}\\
        \begin{minipage}[b]{0.5\hsize}
            \centering
            \includegraphics[width=0.95\hsize]{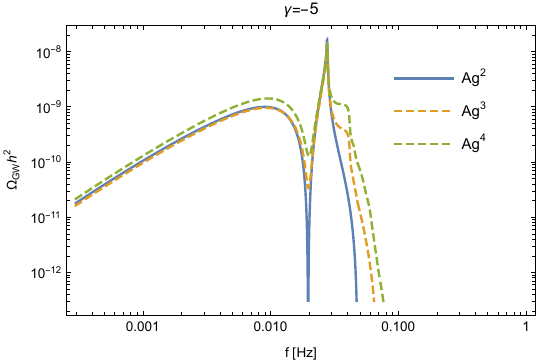}
        \end{minipage}
    \end{tabular}
	\caption{The \ac{SIGW} plots for  $\gamma=-3$ (top left), $\gamma=-4$ (top right), and $\gamma=-5$ (bottom). The blue lines show the contribution order of $\calO(A_g^2)$, the yellow dashed lines for $\calO(A_g^3)$, and the green dashed lines for $\calO(A_g^4)$.  
    }
	\label{fig: GW_conv_negative_gamma}
\end{figure} 

\begin{figure} 
    \centering
    \begin{tabular}{c}
        \begin{minipage}[b]{0.5\hsize}
            \centering
            \includegraphics[width=0.95\hsize]{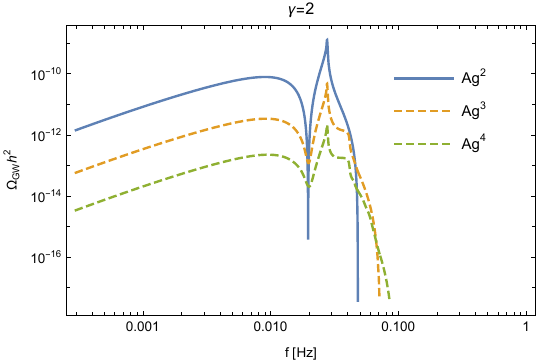}
        \end{minipage}
        \begin{minipage}[b]{0.5\hsize}
            \centering
            \includegraphics[width=0.95\hsize]{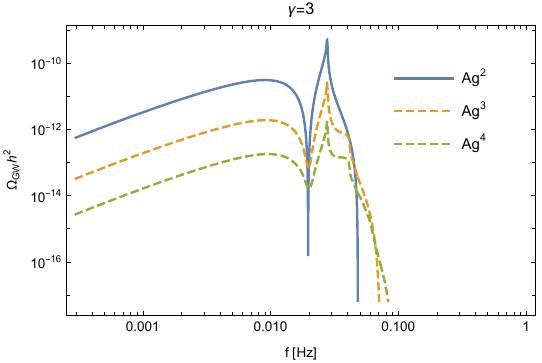}
        \end{minipage}\\
        \begin{minipage}[b]{0.5\hsize}
            \centering
            \includegraphics[width=0.95\hsize]{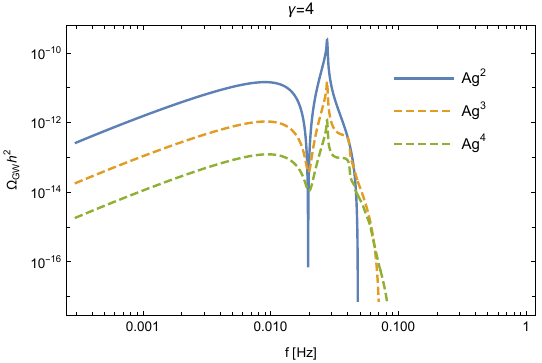}
        \end{minipage}
        \begin{minipage}[b]{0.5\hsize}
            \centering
            \includegraphics[width=0.95\hsize]{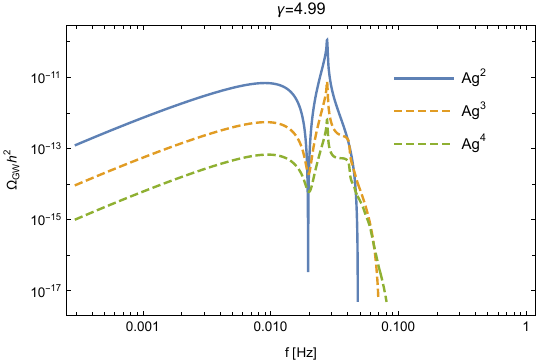}
        \end{minipage}
    \end{tabular}
	\caption{The same plots as Fig.~\ref{fig: GW_conv_negative_gamma} but for $\gamma=2$ (top left), $\gamma=3$ (top right), $\gamma=4$ (bottom left), and $\gamma=4.99$ (bottom right).}
	\label{fig: GW_conv_positive_gamma}
\end{figure}

In this section, we discuss the convergence of the perturbative calculation of \acp{SIGW}
for the negative and positive $\gamma$. The contribution at each order (blue for $\calO(A_g^2)$, yellow dashed for $\calO(A_g^3)$, and green dashed for $\calO(A_g^4)$) for the negative and positive $\gamma$ is shown in Fig.~\ref{fig: GW_conv_negative_gamma} and Fig.~\ref{fig: GW_conv_positive_gamma}. Here, we used $A_g$ values corresponding to $f^{\rm tot}_{\PBH} = 1$ for each plots.
In the negative $\gamma$ region, while the \ac{GW} spectra are well converged for $-4 \lesssim \gamma \leq 0$, the non-Gaussian collections dominate over the leading term in $\gamma \lesssim-4$. Although the perturbative expansion failed for $\gamma \lesssim-4$, we can safely compute the maximum \ac{GW} spectrum around $\gamma = -3.1$ which corresponds to the maximum value of $A_g$ in our model. On the other hand, in the positive $\gamma$ region, the \ac{GW} spectra are well converged and one can safely neglect higher-order corrections.

\section{The BSSN formalism of Numerical Relativity}
\label{appendix:NR_spherical}
In all generality, the line element in the $3+1$ dimension reads
\begin{equation}
\label{timeline_AP}
 \dd{s}^2 = - \alpha^2 \dd{t}^2 + \gamma_{ij}(\dd{x^i} + \beta^i \dd{t})(\dd{x^j} + \beta^j  \dd{t}),
\end{equation}
where $\gamma_{ij}$ is the metric of the 3-dimensional hypersurface, and the lapse and shift gauge parameters are given by $\alpha$ and $\beta^i$ {respectively}. We follow with the conventional  conformal decomposition of the 3-metric, 
$\gamma_{ij} = \psi^4\tilde\gamma_{ij} $
where $\psi$ is the conformal factor. The extrinsic curvature $K_{ij}$ is split into its conformal traceless part $\tilde A_{ij}$ and trace $K$, 
\begin{equation}
K_{ij} = \psi^4 \left( \tilde A_{ij}+\frac13\tilde\gamma_{ij} K\right)~.
\end{equation}
For numerical stability, the first spatial derivatives of the metric $\hat\Delta^i \equiv  - \partial_j\tilde\gamma^{ij}$ are considered as dynamical variables.

In spherical symmetric configurations, the spatial metric element reduces to
\begin{equation}
    \dd{l}^2 = e^{4 \chi(r,t)} \left( a(r,t) \dd{r}^2 + r^2 b(r,t) \dd{\Omega}^2 \right) \; ,
\label{eq:spheremetric}
\end{equation}
where $ a(r,t) \equiv \tilde\gamma_{rr} $ and $b(r,t) \equiv  \tilde\gamma_{\theta\theta}/r$ are coefficients related to the conformal metric, and $\chi (r,t)$ is a redefinition of the conformal factor. We also redefine traceless extrinsic curvature using mixed component $A_a \equiv \gamma^{rr} A_{rr}$ and $A_b \equiv \gamma^{\theta\theta} A_{\theta\theta} = \gamma^{\phi\phi} A_{\phi\phi} $, where naturally the relation $A_a = -2 A_b$ holds. 

The evolution equations of the BSSN dynamical variables are then given by 
\begin{eqnarray}
\partial_t \chi &=& \beta^r \partial_r \chi 
- \frac{1}{6} \alpha K \; , \\
\partial_t a &=& \beta^r \partial_r a + 2 a \partial_r \beta^r 
- 2 \alpha a A_a , \qquad \\
\partial_t b &=& \beta^r \partial_r b + 2 b \: \frac{\beta^r}{r} 
- 2 \alpha b A_b \; , \\
\partial_t K &=& \beta^r \partial_r K - \nabla^2 \alpha
+ \alpha \left( A_a^2 + 2 A_b^2 + \frac{1}{3} \: K^2\right) \nonumber \\
&+& 4 \pi \alpha \left( \rho + S_a + 2 S_b \right) , \\
\partial_t A_a &=& \beta^r \partial_r A_a - \left( \nabla^r \nabla_r \alpha
- \frac{1}{3} \nabla^2 \alpha \right)
+ \alpha \left( R^r_r - \frac{1}{3} R \right) \nonumber \\
&+& \alpha K A_a - \frac{16}3 \pi \alpha \left( S_a - S_b \right) \; , \\
\partial_t \hD^r &=& \beta^r \partial_r \hD^r - \hD^r \partial_r \beta^r
+ \frac{1}{a} \partial^2_r \beta^r + \frac{2}{b} \:
\partial_r \left( \frac{\beta^r}{r} \right)  
\nonumber \\* &-& 
\frac{2}{a} \left( A_a \partial_r \alpha
+ \alpha \partial_r A_a \right)
+ 2 \alpha \left( A_a \hD^r - \frac{2}{rb}
\left( A_a - A_b \right) \right) 
\nonumber \\* &+&   
\frac{2\alpha}{a} \left[ \partial_r A_a
- \frac{2}{3} \: \partial_r K 
+ 6 A_a \partial_r \chi 
+ \left( A_a - A_b \right) \left( \frac{2}{r}
+ \: \frac{\partial_r b}{b} \right) 
- 8 \pi S_r \right] \; ,
\end{eqnarray}
where $\rho$ and $S_r$ are, respectively, the total energy density and the covariant component of the momentum density which are defined below. 
Note that the system of equations respects covariance, only that the terms of $\tilde\gamma^{rr} = a^{-1}$ and $\tilde\gamma^{\theta\theta} = \tilde\gamma^{\phi\phi} = 1/(br^2)$ have been explicitly written. 
The covariant derivatives with respect to the four-metric for the lapse function are computed as follows
\begin{eqnarray}
\nabla^2 \alpha &=&  \frac{1}{a e^{4 \chi}} \left[ \partial_r^2 \alpha
- \partial_r \alpha \left( \frac{\partial_r a}{2a}
- \frac{\partial_r b}{b}
- 2 \partial_r \chi - \frac{2}{r} \right) \right], \\
\nabla^r \nabla_r \alpha &=& \frac{1}{a e^{4 \chi}} \left[ \partial_r^2 \alpha 
- \partial_r \alpha \left( \frac{\partial_r a}{2a}
+ 2 \partial_r \chi \right) \right] \; ,
\end{eqnarray}
and the components of the Ricci tensor and Ricci scalar, in spherical symmetry,  are given by 
\begin{eqnarray}
R^r_r &=& - \frac{1}{a e^{4 \chi}} \left[ \frac{\partial^2_r a}{2a} 
- a \partial_r \hD^r  - \frac{3}{4} \left( \frac{\partial_r a}{a} \right)^2
\right. 
+ \frac{1}{2} \left( \frac{\partial_r b}{b} \right)^2
- \frac{1}{2} \hD^r \partial_r a + \frac{\partial_r a}{rb} \nonumber \\*
&+& \frac{2}{r^2} \left( 1 - \frac{a}{b} \right) \left( 1
+ \frac{r \partial_r b}{b} \right)
+ 4 \left. \partial^2_r \chi - 2 \partial_r \chi \left( \frac{\partial_r a}{a}
- \frac{\partial_r b}{b} - \frac{2}{r} \right) \right] ,
\label{eq:sphere-Rrr} \\
R &=& - \frac{1}{a e^{4 \chi}} \left[ \frac{\partial^2_r a}{2a}
+ \frac{\partial^2_r b}{b} - a \partial_r \hD^r
- \left( \frac{\partial_r a}{a} \right)^2 \right. 
+ \frac{1}{2} \left( \frac{\partial_r b}{b} \right)^2
+ \frac{2}{rb} \left( 3 - \frac{a}{b} \right) \partial_r b \nonumber \\
&+& \frac{4}{r^2} \left( 1 - \frac{a}{b} \right)
+ 8 \left( \partial^2_r \chi + ( \partial_r \chi )^2 \right) 
- \left. 8 \partial_r \chi \left( \frac{\partial_r a}{2a}
- \frac{\partial_r b}{b} - \frac{2}{r} \right) \right] . \hspace{10mm}
\label{eq:sphere-RSCAL}
\end{eqnarray}

\subsection{Gravitational hydrodynamics} 
We consider cosmological settings in the radiation domination epoch where assume the energy-momentum tensor of the perfect fluid, 
\be
T_{\mu\nu} = (\rhofl + p) u_\mu u_\nu + p\> g\munu  ~,
\eeq
where $\rhofl$ and $p$ are the fluid's energy density and pressure, respectively, and $u_\mu$ is the four-velocity. We use a barotropic equation of state, with $p = \omega  \rhofl $ and fix $\omega = 1/3$ according to a radiation fluid. From a comoving observer, the fluid's energy density is commonly described by  splitting it into the rest energy density $\rhozero$ and the internal energy $\varepsilon$, such that 
$\rhofl \equiv \rhozero ( 1 + \varepsilon)$. 
In such a way, for an ideal fluid corresponding to the ultra-relativistic matter (i.e., radiation), we have $\varepsilon \gg 1$, and so $\rhofl \simeq \rhozero \varepsilon$. For an Eulerian observer, the total energy density $\rho$ is then given by 
\be
\rho   = (\rhofl + p) W^2 - p 
~, 
\eeq
where $ W \equiv 1 / {\sqrt{ 1-v^2 }}$ is the Lorentz factor, with $v^i$ being the three-velocity component of the fluid as seen by the Eulerian observer.  

The state of the fluid is described at any time by the \textit{primitive variables} $\rhozero$, $\varepsilon$, $p$, and $v^i$. The evolution equations of such a system can be derived from the conservation laws 
\be
 \nabla_\mu \left(  \rhozero u^\mu \right) = \partial_\mu \left(  \sqrt{-g}\> \rhozero u^\mu  \right)  = 0 ~,
\eeq
and
\be
\nabla_\mu T^\mu_\nu  = \partial_\mu \left( \sqrt{-g}\> T^\mu_\nu \right)  - \sqrt{-g}\> \Gamma^\alpha\munu  T^\mu_\alpha  = 0 ~. 
\eeq
In here, we skip the full derivation of the equations of motion as these can be found in full detail in Ref.~\cite{10.1093/acprof:oso/9780199205677.001.0001}. 
The procedure leads to a system of equations where a new set of variables is defined to be used during the evolution. These are the \textit{conserved variables} $D$, $S_i$ and $\mathcal{E}$,  and they are defined in terms of the primitive variables, such that
\begin{align}
 D &\equiv \rhozero W  ~,  \label{eq:D}
\\
 S_i &\equiv  (\rhofl + p) W^2 v_i  ~,   \label{eq:Si}  %
\\
\mathcal{E} &\equiv  (\rhofl + p) W^2 -p - D     \label{eq:calE}   ~.  
\end{align}
The evolution equation for the fluid, written in terms of BSSN variables, is then given by  
\begin{align}
\left( \partial_t -  \mathcal{L}_\beta \right) D  & =   - D_k (\alpha D v^k) + \alpha K D  ~,   \label{eq:evoD}
\\
 \left( \partial_t - \mathcal{L}_\beta \right) S^i  &=    - D_k \left[ \alpha  \left( S^i v^k  + \gamma^{ik} p \right) \right]   \nonumber \\ & \> - (\mathcal{E} + D) D^i\alpha  + \alpha K S^i  ~,     \label{eq:evoSi}
\\
\left( \partial_t - \mathcal{L}_\beta \right) \mathcal{E} & =
 (\mathcal{E} + D + p) (\alpha v^mv^n K_{mn} - v^m \partial_m \alpha) \nonumber \\
 & \>  - D_k \left[ \alpha v^k  \left( \mathcal{E} + D \right) \right]   + \alpha K (\mathcal{E} + p) ~,  \label{eq:evoE}
\end{align}
where $D_i$ denotes the covariant derivative with respect to the spatial metric. 
A generic equation of state often requires a root-finding technique to obtain the value of $p$ and $v^i$, prior to the recovery of the other variables, as they are needed to evolve Eqs.~\eqref{eq:evoD}--\eqref{eq:evoE}. However, for cases with a simple equation of state, the recovery of $p$ and $v^i$ can be done analytically by finding the physical root of a high-order polynomial. For a barotropic fluid with $p = \omega \rhofl $, this implies solving a second-order polynomial \cite{Staelens:2019sza}, which diminishes the computational burden of the simulations.


\bibliography{main}
\bibliographystyle{JHEP_mod}
\end{document}